\documentclass[aps,prl,twocolumn,showpacs,groupedaddress]{revtex4}

\newcommand{\eqb}{\begin{equation}}
\newcommand{\eqe}{\end{equation}}
\newcommand{\arrb}{\begin{eqnarray}}
\newcommand{\arre}{\end{eqnarray}}

\usepackage{graphicx}
\usepackage{subfigure}
\usepackage{dcolumn}

\begin{document}
\title{Saturated Hydrocarbons on Silicon: Quantifying Desorption with
Scanning Tunneling Microscopy and Quantum Theory}
\author{N. L. Yoder, N. P. Guisinger, and M. C. Hersam}\email{m-hersam@northwestern.edu}
\affiliation{Department of Materials Science and Engineering,
Northwestern University, 2220 Campus Drive, Evanston, IL 60208-3108}
\author{R. Jorn, C.-C. Kaun, and T. Seideman}\email{t-seideman@northwestern.edu}
\affiliation{Department of Chemistry, Northwestern University, 2145
Sheridan Road, Evanston, IL 60208-3113}
\begin{abstract}
Electron stimulated desorption of cyclopentene from the Si(100)-2x1
surface is studied experimentally with cryogenic UHV STM and
theoretically with transport, electronic structure, and dynamical
calculations. Unexpectedly for a saturated hydrocarbon on silicon,
desorption is observed at bias magnitudes as low as 2.5 V, albeit
the desorption yields are a factor of 500 to 1000 lower than
previously reported for unsaturated molecules on silicon. The low
threshold voltage for desorption can be attributed to hybridization
of the molecule with the silicon surface, which results in low-lying
ionic resonances within 2-3 eV of the Fermi level. These resonances
are long-lived, spatially localized and displaced in equilibrium
with respect to the neutral state, resulting, upon excitation, in
symmetric (positive ion) or asymmetric (negative ion) motion of the
silicon dimer atoms. This study highlights the importance of nuclear
dynamics in silicon-based molecular electronics and suggests new
guidelines for the control of such dynamics.
\end{abstract}
\maketitle

The scanning tunneling microscope (STM) has
progressed beyond basic imaging of surfaces to become a tool for
manipulating single atoms and molecules. Examples include reversible
vertical switching \cite{Eigler90}, lateral displacement
\cite{Bartels97,Lastapis05,Basu05}, rotation \cite{Ho02},
dissociation \cite{Ho02}, and desorption
\cite{Shen95,Hersam00,AlaviPRL00}. These new experimental
techniques, combined with related theoretical
studies \cite{Seideman03,AlaviPRL00,AlaviJCP00}, have enabled
detailed understanding of fundamental physical processes at
interfaces. In the emerging field of molecular electronics
\cite{Aviram74,Joachim00,Nitzan2003,Tour2000,Mantooth2003,Kushmerick2004}, significant experimental
and theoretical effort has been invested into studying the
electronic properties of single molecules, molecular nanostructures,
and organic monolayers. A subfield of increasing interest is the
integration of molecular electronic devices with conventional
silicon microelectronic technology \cite{Blum2005,Beckman2006,Lenfant2003,Wolkow1999,Hersam04,Rakshit04}. A
detailed understanding of both the electronic properties and the
stability of organic molecules on semiconductors must be
established, however, before reliable devices can be realized. On
the one hand, it is relevant to develop general guidelines for
molecules that will provide stable devices. In particular, it is
widely believed that saturated organic/silicon systems offer
stability with respect to current-induced failure of silicon-based
molecular electronics \cite{Patitsas00}. It is thus of both
fundamental and practical importance to explore the extent to which
this expectation holds true. On the other hand, it is desirable to
explore systems that will provide dynamical function, such as
current-driven molecular machines \cite{Seideman03}, switches or
rectifiers \cite{Aviram74}. Previous work has illustrated the
suitability of STM experiments, combined with structural and
dynamical calculations, to address both questions
\cite{Seideman03,AlaviPRL00}.

In this Letter, we study the desorption of cyclopentene from a
silicon surface using the combination of STM measurements,
electronic structure theory, quantum transport, and dynamical
calculations. The efficiency of the desorption process is
characterized by the yield, which is found to be a factor of
500--1000 lower than for benzene/Si(100)
\cite{AlaviPRL00,AlaviJCP00} or chlorobenzene/Si(111) - 7x7
\cite{Sloan03}. Measurements of the desorption yield for sample bias
voltages from -5 V to +5 V show a clear turn-on behavior, with
asymmetric threshold voltages. Yield measurements as a function of
tunneling current strongly suggest that the desorption mechanism is
a single electron process, resulting from transient excitation of a
resonant state. Quantum transport theory, along with electronic
structure and dynamical calculations give insight into the nature of
the resonances and the associated nuclear dynamics.

Experiments were performed using a cryogenic, variable temperature 
ultra-high vacuum (UHV)
STM \cite{Foley04} operating at 80 K, and heavily boron doped
Si(100) substrates ($\rho$ $<$ 0.005 $\Omega$-cm). Details of the
experimental procedure for producing a submonolayer coverage of
cyclopentene on the clean Si(100)-2x1 surface are given elsewhere
\cite{Guisinger05}. A quantitative measure of the desorption yield
was obtained within a 3-step statistical approach. The surface is
initially imaged at non-perturbative conditions (-2 V, 0.1 nA) 
to determine the initial number of molecules on the surface.
The tip is then scanned over the same area at elevated sample bias
and/or tunneling current conditions in order to induce desorption.
Finally, the same area is scanned a third time at -2 V, 0.1 nA, and
the number of desorbed molecules is counted. The yield
(events/electron) is given by Y = N$_{D}$*(e/I*t), where N$_{D}$ is
the number of desorption events, I is the tunneling current, e is
the electronic charge, and t is the total time the tip spent over a
molecule. Overcounting of the electrons (since the current continues
to flow after desorption occurs) is corrected using first-order
kinetics \cite{Trenhaile05}.

The electronic transport was computed ab initio within a
non-equilibrium Green's function approach \cite{Jauho94}, using a
density functional theory (DFT) Hamiltonian expanded in a real-space
basis with atomic cores defined by pseudopotentials. This
formulation allows for proper description of the electronic
properties through the use of an extended (open boundary)
representation of both the tip and the substrate, extending to
infinity in the vertical direction and, for the surface, including
periodic boundary conditions in the lateral directions. The
electronic structure calculations were carried out using DFT with
the B3LYP functional and both 6-31G* and 6-311G**
basis sets within Q-Chem \cite{Kong00}.
\begin{figure}
\begin{center}
\resizebox{8.5cm}{!}{\includegraphics{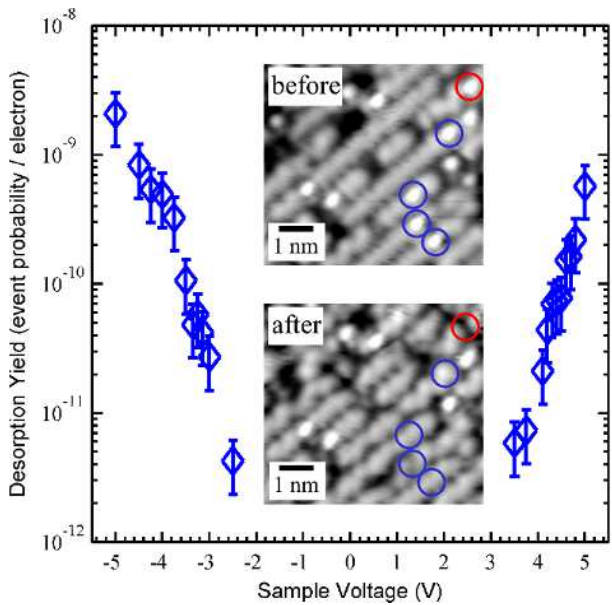}}
\end{center}
\caption{(color online). Desorption yield as a function of the sample bias
between -5 V and 5 V (I = 2 nA). (inset) STM images of cyclopentene
on Si(100)-2x1 before and after desorption. The desorption of
cyclopentene usually leaves behind a clean silicon dimer (blue
circles), but occasionally results in an apparent depression in
place of the molecule (a red circle). For the purposes of this
study, we have treated both cases as a desorption event. }
\end{figure}

The adsorption of cyclopentene on silicon substrates has been
studied experimentally \cite{Hamers97,Yamashita02} and
computationally \cite{Cho01}, and has been explained by a [2 + 2]
cycloaddition reaction between the C=C bond of the cyclopentene and
a silicon dimer, resulting in the formation of two covalent Si-C
bonds. In STM images, cyclopentene molecules appear as protrusions
centered on the silicon dimer row, as shown in Fig. 1 (inset). The
desorption yield given in Fig. 1 shows that desorption occurs over a
wide range of negative and positive bias voltages, and is strongly
voltage-dependent, with the yield increasing exponentially after a
threshold voltage is reached. A yield of 1x10$^{-12}$ is the
practical lower-limit for desorption measurements and is thus used
to define the threshold voltage. We find asymmetric behavior,
with a threshold voltage of -2.5 V at negative, and 3.5 V at
positive sample bias. The dramatic sensitivity of the desorption
yield to the sample bias rules out the possibility of an electric
field-induced desorption (often difficult to distinguish from other
mechanisms in STM experiments), suggesting an electronic mechanism.
Measurements of the desorption rate (determined as the ratio of the
desorption yield and the tunneling current) versus the tunneling current
found very weak dependence of the rate on the current in the 0.1--2
nA range at -5 V sample bias. This approximately linear dependence
rules out the possibility of a ``vibrational heating''
multiple-electron process, indicating that the desorption of
cyclopentene occurs via a single-electron resonant process.

Several previous desorption studies
\cite{AlaviPRL00,Sloan03,AlaviJCP00} of organic molecules on silicon
report threshold voltages (1-3 V) similar to those reported here,
and describe the desorption mechanism in terms of a charge carrier
resonance scattering event. The presence of $\pi$-character in
organic molecules bound to silicon gives rise to low-lying ionic
resonances within a few eV of the Fermi level. Scattering via these
resonances transiently promotes the molecule to an ionic potential
energy surface on which the nuclei evolve subject to a finite
lifetime. Provided that the equilibrium configurations of the
resonance and neutral states differ appreciably, the molecule will
attempt to relax while in the charged state, resulting in dynamics
on the ionic surface \cite{Seideman03}. As a result, subsequent to
the resonance tunneling event, the molecule can be placed in an
excited configuration on the neutral surface, resulting in further
dynamics such as bond breaking and desorption.

Cyclopentene is believed to be stable toward such resonance
processes since it does not retain $\pi$-orbital character upon
chemisorption to the clean Si(100)-2x1 surface, and is therefore
expected to lack resonances sufficiently close to the Fermi level to
trigger desorption. The observation of desorption at both bias
polarities over a wide range of sample biases and tunneling currents
suggests the existence of a new avenue for resonance-mediated
dynamics, which is of fundamental interest and may also be of
technological relevance to the search for stable constructs. The
desorption yield of cyclopentene from Si(100), ranging from
2x10$^{-12}$ to 1x10$^{-8}$ events/electron, is between 500 and
1000 times smaller than the reported yields for benzene on
Si(100)-2x1 \cite{AlaviPRL00} or chlorobenzene on Si(111)-7x7
\cite{Sloan03},  suggesting significant differences in the
desorption dynamics following resonant excitation.

To understand the desorption mechanism and explore its generality
and implications, we applied a combination of ab-initio transport,
structural, and dynamical calculations within the formalism of
\cite{Seideman03}. We began by investigating the electronic
properties of the underlying resonances using the NEGF-DFT method
described above. To compute the energetic location of the resonance states and
explore their nature, we first diagonalized the sub-Hamiltonian
corresponding to the atomic orbitals of the cyclopentene molecule to
obtain the respective eigenvalues and eigenfunctions. The highest
occupied molecular orbital (HOMO) and lowest unoccupied molecular
orbital (LUMO) of this system are located at -2.49 eV and 6.90 eV,
respectively, with respect to the Fermi level of the system. We then
calculated the eigenvalues and eigenfuctions of the sub-Hamiltonian
matrix corresponding to the cyclopentene and the two attached Si
dimer atoms. For this system, the HOMO and LUMO are located at -2.00
eV and 2.95 eV, respectively, signifying strong hybridization of the
molecule with the substrate. The signature of these states can also
be seen in Fig. 2, which shows the projected density of states (PDOS)
for the (a) HOMO and (b) LUMO.  Importantly, the sharp maximum in
the PDOS in Fig. 2 correspond to maxima in the computed transmission
coefficient, showing that these sharp resonances play a major role
in the conductance and hence also in the current-induced dynamics.

These results indicate that strong hybridization of the molecule
with the silicon substrate introduces new states into the HOMO-LUMO
gap of cyclopentene, which are accessible at rather low bias
voltages. In contrast with the resonances in unsaturated systems,
where the electronic orbitals are spatially delocalized and the
equilibrium displacements with respect to the neutral state are
generally small, the electronic orbitals in saturated molecules are
typically localized, giving rise to substantial equilibrium
displacements.
\begin{figure}
\begin{center}
\resizebox{8.5cm}{!}{\includegraphics{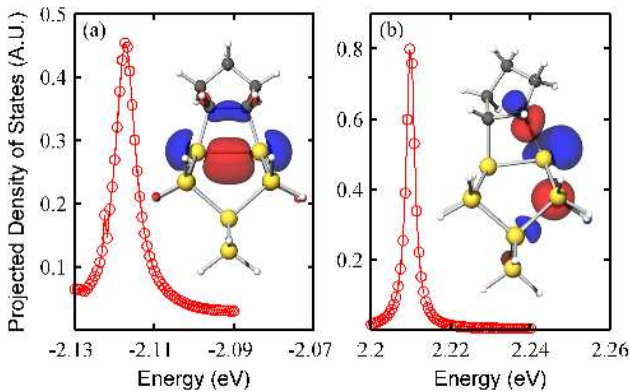}}
\end{center}
\caption{(color online). (a) Peak in PDOS corresponding to the positive
ion. Positive Ion HOMO plot (inset) shows symmetric localization of the
charge primarily on the silicon dimer. (b) Peak in PDOS
corresponding to the negative ion. Inset shows HOMO of the negative
ion, showing significant localization of the electron onto the `up'
silicon dimer atom, with some population on the attached C and Si
atoms. } \label{orbitalfig}
\end{figure}
The localized nature of the molecular orbitals of
cyclopentene/Si(100) can be seen in Fig. 2 (insets), where we show
the results of electronic structure calculations of the positive and
negative ion systems within a cyclopentene-Si$_{9}$H$_{12}$ model.
The HOMOs of both ionic states are plotted in order to visualize the
spatial extent of the charge distributions.

The importance of the hybridization of the molecule with the silicon
dimer in forming the ionic resonances was further supported by
performing L\"{o}wdin and natural 
population analyses (LPA and NPA, respectively). Differences between
the charges residing on atoms in the neutral and ionic states
provide insight into the spatial localization of the
resonances. For the cationic state, the atomic charge increases
equally on both Si dimer atoms, yielding an increase from the
neutral state charges of +0.15e within the LPA and +0.31e within the
NPA analysis. In the case of the negative ion state, the difference
in charge residing on the raised silicon of the tilted dimer is
-0.18e within LPA and -0.27e within NPA. The population analysis
confirms that the ionic states display a substantial localization of
the charge around the silicon dimer as compared to the neutral state
and reinforces the NEGF observation that the silicon dimer plays a
crucial role in the formation of the resonances. Additionally, while
the localization of the charge in the positive ion is symmetric with
respect to the two Si dimer atoms, the negative ion case shows an
asymmetric localization of the charge primarily on the raised Si
atom, in agreement with the results displayed in Fig. 2.

To complete our analysis of the resonance states and understand the
low desorption yield observed, we computed the lifetimes of the
ionic resonances. In the case of isolated resonances, relevant here,
these can be extracted from fitting the PDOS to a Lorentzian form.
The obtained lifetimes of the HOMO and LUMO are 94 fs and 257 fs,
respectively. These long lifetimes allow for significant evolution
of the vibrational modes in the ionic state and hence substantial
transfer of electronic energy into vibrational excitation in the
course of the tunneling event, enabling desorption. At the same
time, the long lifetime implies that only a small fraction of the
current is mediated via the resonance features. Expressing the
desorption rate as an (energy integrated) product of a desorption
probability per resonance event multiplied by the rate of electron
transport through the resonance \cite{Seideman03}, one finds that
the narrow features translate into low reaction rates, consistent
with our observations.
\begin{figure}
\begin{center}
\resizebox{8.5cm}{!}{\includegraphics{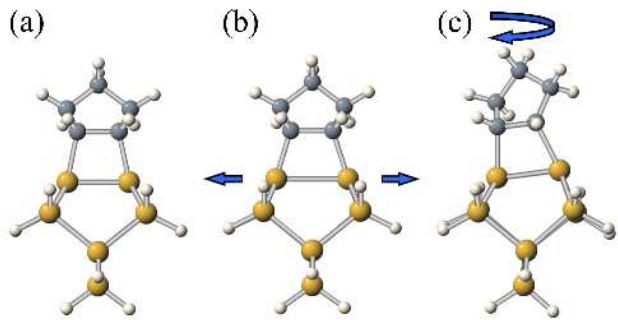}}
\end{center}
\caption{(color online). Optimized equilibrium geometries of cyclopentene
on a Si$_{9}$H$_{12}$ cluster: (a) neutral molecule geometry, (b)
positive ion geometry, and (c) negative ion geometry.}
\label{dftgeometries}
\end{figure}

In order to investigate the nature of the nuclear motion following
the charge transfer event, we proceeded with a DFT study of the
reaction dynamics within a cyclopentene-Si$_{9}$H$_{12}$ model.
Previous work \cite{AlaviJCP00} has shown that, for the
electronically localized case of a small organic molecule on
Si(100), a cluster of this size suffices to capture the physics
while being computationally accessible. Interestingly, as shown in Fig. 3, 
both ionic equilibrium geometries
are markedly displaced with respect to the neutral geometry, with
the positive ion showing a ($\sim$15$\%$) stretch of the Si-Si dimer
bond, while the negative ion exhibits significant buckling of the Si
dimer as well as twisting of the cyclopentene ring. By comparison to
other systems where (photon- or electron-triggered) desorption
induced by electronic transitions was observed, cyclopentene/Si(100)
exhibits an order of magnitude longer lifetimes and significantly
more pronounced rearrangement upon charge transport, in particular
in the LUMO case. These large displacements from the neutral state
equilibrium along with the long lifetimes, introduce the possibility
for substantial motion in the resonance state, which we proceeded to
explore. Extensive dynamic reaction coordinate (DRC) simulations in
GAMESS \cite{Schmidt93} confirm that the motion of the molecule
following the scattering event in either charged state does not
directly excite the Si-C bond. Rather, vibration of the Si-Si dimer
bond is excited in the case of the positive ion, while buckling of
the dimer is observed in the negative ion case, consistent with the
ionic equilibrium geometry calculations. In contrast with the
previous reported case of desorption of benzene from Si(100)
\cite{AlaviPRL00,AlaviJCP00}, the initially excited modes are weakly
coupled to the reaction coordinate, resulting in relatively low
desorption probability per resonant event. This result is consistent
with our observation that the desorption yields for cyclopentene are a
factor of 500--1000 smaller than those reported for benzene.

In summary, the combination of cryogenic ultra-high vacuum STM
measurements, with ab-initio transport, electronic structure, and
dynamical calculations illustrate the possibility of desorbing fully
saturated organic molecules from Si(100) at modest sample biases and
provide a detailed picture of the desorption mechanism. Desorption
occurs via resonance attachment of a hole/electron to the HOMO/LUMO
of the substrate-adsorbate system. A key to the feasibility of such
desorption events is strong hybridization of the molecule with the
silicon substrate, which gives rise to new orbitals that are
energetically remote from, and bear no resemblance to the orbitals of 
the isolated molecule. Such previously unexplored desorption pathways
carry interesting implications both to the development of guidelines
for the construction of stable (current-immune) silicon-based
molecular electronics, and to the prospect of designing
current-driven molecular devices.
\begin{acknowledgments}
This work was supported by the National Science Foundation (Grant Nos. 
DMR-0520513, CHE-0313638/002, and ECS-0506802) and the Office of Naval 
Research (Grant No. N00014-05-1-0563). N.L.Y. acknowledges an NSF 
Graduate Research Fellowship and thanks Dr. Eric Brown for useful 
discussions.
\end{acknowledgments}
\bibliography{yoder}
\end{document}